\documentclass[aps,prl,twocolumn,superscriptaddress, 10pt]{revtex4-1}
\usepackage{graphicx} 
\usepackage{xcolor}
\usepackage{braket}
\usepackage{amsmath}

\begin{document}

\title{Quantum Reorientational Excitations in the Raman Spectrum of Hydrogen} 

\author{Philip Dalladay-Simpson}
\email{philip.dalladay-simpson@hpstar.ac.cn}
\affiliation{Center for High Pressure Science and Technology Advanced Research, 1690 Cailun Road, Shanghai, 201203, China}

\author{Eric Edmund}
\affiliation{Center for High Pressure Science and Technology Advanced Research, 1690 Cailun Road, Shanghai, 201203, China}

\author{Huixin Hu}
\affiliation{Center for High Pressure Science and Technology Advanced Research, 1690 Cailun Road, Shanghai, 201203, China}

\author{Mario Santoro}
\email{santoro@lens.unifi.it}
\affiliation{Istituto Nazionale di Ottica, Consiglio Nazionale delle Ricerche (CNR-INO), via N. Carrara 1, 50019 Sesto Fiorentino, Italy}
\affiliation{European Laboratory for Non Linear Spectroscopy (LENS), via N. Carrara 1, 50019 Sesto Fiorentino, Italy}

\author{Federico Aiace Gorelli}
\email{federico.gorelli@sharps.ac.cn}
\affiliation{Center for High Pressure Science and Technology Advanced Research, 1690 Cailun Road, Shanghai, 201203, China}
\affiliation{Istituto Nazionale di Ottica, Consiglio Nazionale delle Ricerche (CNR-INO), via N. Carrara 1, 50019 Sesto Fiorentino, Italy}
\affiliation{European Laboratory for Non Linear Spectroscopy (LENS), via N. Carrara 1, 50019 Sesto Fiorentino, Italy}
\affiliation{Shanghai Advanced Research in Physical Sciences (SHARPS), Shanghai, China}


\begin{abstract}
Low-frequency Raman peaks, below 250 cm$^{-1}$, are observed in hydrogen between 2–174~GPa and 13–300~K. The origin of these features is attributed to reorientational transitions ($\Delta J=0$; $Q_0$-branch), which shift from the Rayleigh line as anisotropic intermolecular interactions lift the $m_J$ degeneracy. This family of excitations closely follows the behavior of the $S_0$-branches, sharing their dependence on pressure, temperature, and ortho-H$_2$ concentration. Above 65 K, spectra corrected by the Bose-Einstein population factor reveal a broad continuum arising from populated higher $J$-states and increased ortho–para disorder. Upon entering phase III, where molecular rotation is inhibited, this continuum is quenched, giving way to well-established optical phonons. Below 25~K, equilibrated samples demonstrate a fine structure from isolated and pair excitations from impurity ortho-H$_2$ molecules in a parahydrogen lattice, the latter a sensitive probe to anisotropic intermolecular interactions relevant to the quantum modeling of solid H$_2$.
\end{abstract}

\pacs{}

\maketitle 

Hydrogen is a prototypical quantum solid, evidenced by its inflated volume due to large zero-point motion, the preservation of its rotational dynamics to 0~K and the existence of nuclear spin isomers \cite{Silvera1980, VanKranendonk1983, Mao1994}. The latter is a consequence of the total nuclear spin state of the molecule, $I$, and the rotational quantum number, $J$, being coupled and results in two distinct molecular species of hydrogen: para-H$_2$ ($I=0$; singlet-state: $\frac{1}{\sqrt{2}}(\ket{\uparrow\downarrow}-\ket{\downarrow\uparrow})$ only occupying even-$J$ rotational levels and ortho-H$_2$ ($I=1$; triplet-state: $\ket{\uparrow\uparrow}$, $\ket{\downarrow\downarrow}$ and $\frac{1}{\sqrt{2}}(\ket{\uparrow\downarrow}+\ket{\downarrow\uparrow})$ only occupying odd-$J$ levels \cite{Dennison1927}. Raman spectroscopy has been foundational in probing this quantum nature since its inception almost a century ago, providing early validation of quantum mechanics \cite{Heitler1927, Dennison1927} through the characterization of the relative intensities of the rotational $S_1(J)$-branch, a direct signature of spin isomers and their thermal equilibrium concentration (1:3, para:ortho) at room temperature \cite{McLennan1929}.

Raman spectroscopy remains pivotal, having mapped out most of hydrogen’s condensed phases prior to its long-sought density-driven metallization \cite{Wigner1935}. This includes the discovery of the broken-symmetry phase (Phase II)\cite{Silvera1981}, the orientationally ordered phase (Phase III) \cite{Hemley1988}, and the layered mixed-molecular phases (Phases IV/V) \cite{Howie2012, Dalladay-Simpson2016, Eremets2011}, providing key constraints for the most advanced quantum models of hydrogen  \cite{Monacelli2023, Monserrat2016, Drummond2015}. Up to 150 GPa, hydrogen's phase diagram is dominated by phase I, which is characterized by a hexagonal close-packed (hcp) lattice of coupled quantum rotors \cite{Silvera1980, VanKranendonk1983, Mao1994}. In phase I, increased intermolecular interactions reduce $J$ as an effective quantum number \cite{Meier2020, Pena-Alvarez2020}. However, where $J$ remains a good quantum number, below $\sim$70 GPa \cite{Meier2020}, the Raman spectrum of phase I is remarkably simple and characterized by vibrational ($\nu_{1\rightarrow0}$), rotational ($S_0(J)$; $\Delta J=2$), and $E_{2g}$ phonon modes. Notably, with the lowest rotational excitation $S_0(0)$ at $~$350 cm$^{-1}$ and the lowest $E_{2g}$ phonon frequency recorded being $~$37 cm$^{-1}$ \cite{Silvera1972}, one might expect a spectral gap below which no Raman excitations are observable.

Anisotropic interactions in hydrogen lift the $2J+1$ degeneracy of the orientational sub-levels, $m_J$, giving rise to low-energy Raman-active reorientational excitations where the $J$ number is preserved, ie. a $Q$-branch. These were first identified over 50 years ago by pioneering Raman spectroscopy measurements on solid hydrogen at ambient pressure and low temperatures\cite{Silvera1972}. Subsequent studies, under similar conditions, detected them in a dilute-ortho-H$_2$ solid via microwave absorption \cite{Hardy1975} and, later, as combination bands with infrared absorption \cite{Dickson1996}. The excitations are characteristically very low in Raman frequency, less than 6~cm$^{-1}$ \cite{Hardy1977}, making them particularly challenging to observe given their proximity to the Rayleigh line. However, high pressure is expected to amplify the underlying anisotropic interactions, thereby increasing their associated frequencies and intensities into more readily experimentally accessible regimes. Recent simulations have suggested that this reorientational branch exhibits a fine structure at room temperature with components reaching frequencies $\sim$7-70~cm$^{-1}$ at 40~GPa \cite{Cooke2020}. However, because of their symmetry, they are not observable in a perfectly back-scattering geometry with the c-axis parallel to the collection direction \cite{Cooke2020, Pena-Alvarez2020}. Therefore, due to the crystallographic preferred orientation of hydrogen in a Diamond Anvil Cell (DAC) \cite{Loubeyre1996, Akahama2017} this has remained an experimental blind spot for high-pressure studies. As a consequence, their only tentative identification in D$_2$ required a 90$^{\circ}$ scattering geometry \cite{Silvera1972}, a technical limitation for high-pressure experiments.

In this study low-frequency Raman peaks, below 250~cm$^{-1}$, have been observed between 2-174~GPa and 13-300~K. We interpret the origin of these features as reorientational transitions, a $Q_0$-branch, between orientational sub-levels ($m_J$) within the same rotational state $J$ of the ground vibrational manifold. The appearance of these transitions is due to the lifting of $m_J$ degeneracy, caused by anisotropic nearest neighbor ($nn$) interactions in the crystalline lattice, without which they would occur at zero-frequency. The Bose-Einstein population (BEP) corrected spectral line shape is found to be highly dependent on temperature and ortho-H$_2$ concentration. At higher temperatures, above 65~K, a low-frequency broad Boltzmann-like distribution is observed, consistent with a thermal population of higher $J$-states and increased ortho-para $nn$ configurational disorder. These excitation's behavior is similar to that of the rotational $S_0$-branches, further supporting the interpretation that both arise from quantum orientational motion. Upon entering phase III, where molecular rotation is arrested, this broad distribution quenches into the well-resolved, known phonons of that phase \cite{Goncharov2001_PNAS, Goncharov1998}. In phase I at low temperatures, below 25~K, and dilute ortho-concentrations, less than 5~\%, a resolved fine structure emerges, indicative of impurity pair and isolated ortho-H$_2$ excitations \cite{VanKranendonk1983, Hardy1975, Silvera1980}. This pressure-dependent fine structure serves as a sensitive probe of the anisotropic intermolecular interactions essential for constraining models of solid hydrogen.

Figs. 1(a, c), 2 and End Matter Fig. 5(a, b) present raw measurements on samples of hydrogen in phase I at various pressures and down to 17~K, these spectra reveal the expected rotational $S_0$-branches of a quantum rotor and the weaker optical $E_{2g}$ phonon of the hcp lattice. Across all spectra above 2.4~GPa, we observe a prominent and previously unaccounted-for feature characterized by a strong increase in intensity toward zero frequency in both the Stokes and anti-Stokes regions. As a result, this unusual lineshape appears almost like a peak centered at zero frequency, which is non-physical for a Raman excitation. As a consequence, this feature could easily be mistaken for an experimental artifact, such as poor filtration of the Raman excitation source and/or sample/diamond photoluminescence and perhaps explains why it has been disregarded in previous studies \cite{Wang2019, Pena-Alvarez2020}. 

\begin{figure}
    \centering
    \includegraphics{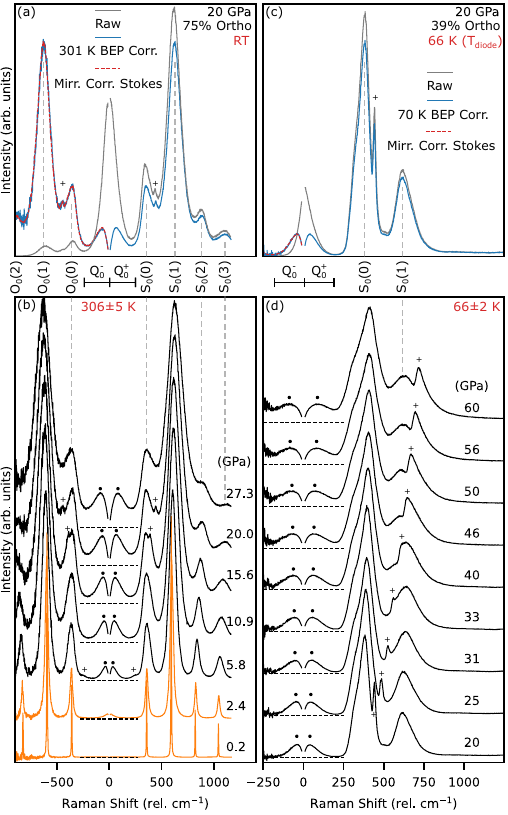}
    \caption{532~nm excited Raman spectroscopy of H$_2$ during isothermal compressions; (1) left panels (a,b), room temperature from 0.2-27.3~GPa and (2) right panels (c,d) at 66~K from 20-60 GPa. Fluid H$_2$ and solid H$_2$ are plotted with orange and black traces respectively. Top panels (a) and (c) present a raw Raman spectrum at 20~GPa (gray), the BEP corrected spectrum (blue) and the mirrored corrected Stokes spectrum (dashed red). Bottom panels (b) and (d), are waterfall plots of the corrected spectra, with the average temperature and its error bar (one standard deviation (SD)) determined spectroscopically from the BEP correction. Dashed horizontal lines mark the zero-intensity baselines for each spectrum. Dots and crosses denote the maxima of the $Q_0$-branches and the frequency of the $E_{2g}$ phonon peak, respectively. The rotational excitations are highlighted as $S_0(J)$ and $O_0(J)$, for Stokes and anti-Stokes, respectively.}
    \label{Fig1}
\end{figure}

For low-frequency excitations at finite temperatures, the Raman lineshape is significantly influenced by the Bose-Einstein population factor, which can result in distorted profiles \cite{Kauffmann2019, Dalladay-Simpson2025}. To recover the bare Raman cross-section, we applied a temperature-fitted BEP correction, dividing the Stokes and anti-Stokes intensities by $(1+n(\omega,T))$ and $n(\omega,T)$, respectively, where $n(\omega,T)$ is the Bose-Einstein factor (See End Matter for details). The resulting BEP-corrected spectrum, which is symmetric around zero-frequency, is shown by the blue trace in Figs.~\ref{Fig1}(a, c) and Fig. 5(a) as well as the waterfall of spectra in Figs. 1(b, c), 3, 5(b, c) and 6. This correction successfully resolves the strong peak centered around the Rayleigh line into a weaker but distinct low-energy Stokes/anti-Stokes pair of excitations, with Boltzmann-like distributions up to 250~cm$^{-1}$. The validity of this procedure is confirmed by the excellent agreement between the anti-Stokes side of the BEP-corrected spectrum and the mirrored corrected Stokes side, as shown by the red dashed traces in Figs.~\ref{Fig1}(a, c) and 5(a). This agreement unambiguously validates the signal as symmetric, first-order Raman activity around the Rayleigh line as seen in Figs. 1, 5 and 6. It is also further substantiated by being observed with both 532 and 660~nm Raman excitation wavelengths, shown in Figs. 1 and 5, respectively. Interestingly, despite hydrogen's transparency to visible light, a marginal laser heating effect is evident in all measurements, with fitted spectral temperatures up to 10~K higher than those measured on the DAC body.  

Given the low frequency of these excitations, with the tail of the distribution approaching zero-frequency and therefore zero energy, we assign it to a family of low-energy reorientational transitions \cite{Hardy1977, Silvera1980, Silvera1972}. These excitations, generated by transitions between the $m_J$ orientational sub-levels, occur within the vibrational ground state and preserve the initial rotational quantum number $J$ (for $J>0$). For an isolated molecule, the $m_J$ sub-levels are 2$J$+1-fold degenerate, but in a condensed state, anisotropic intermolecular interactions partially split them, shifting the transition energy away from zero \cite{Silvera1972, Silvera1980, Hardy1975, Hardy1977, VanKranendonk1983}. From this point forward, we will refer to these quantum reorientational excitations as a $Q_0$-branch ($Q_0^+$ for Stokes, $Q_0^-$ for anti-Stokes), the spectroscopic notation assigned due to the preservation of $J$-state for the Raman transition. As mentioned previously, these branches were recently predicted to contribute to the Raman spectrum at high pressure \cite{Cooke2020}, but was suggested to be unmeasurable in a back-scattering geometry \cite{Cooke2020, Pena-Alvarez2020}. We attribute our successful observation to the high numerical aperture (NA) of the objective used, partially mitigating the orientational dependence of these excitations \cite{Cooke2020, Pena-Alvarez2020}. Given at ambient pressure and low temperature a fine structure  is characteristic for these excitations, as reported in both microwave absorption and infrared absorption measurements \cite{Hardy1975, Dickson1996}. We suggest that at elevated temperature, the fine-structure associated with orientational transitions is smeared out by ortho-para $nn$ configurational disorder and thermal population of higher $J$-states, a phenomenon which is well-established for the rotational $S_0$-branches \cite{Goncharov2001_PRB, Silvera1980}.

The behavior of this peak under pressure and temperature closely mirrors that of the $S_0$ rotational branches, as observed in both the BEP-corrected and uncorrected spectra (Figs. 1(b), 1(d), and 5). As seen in Fig. 1(b), the $S_0$-branches are distinctly sharp in the low-density fluid state at 0.2~GPa, $<$5~cm$^{-1}$ FWHM, and as a result the $Q_0$-branch is not resolved from the Rayleigh line. However, above 2.4~GPa, whilst still in a fluid-state, a small peak is found to emerge as the rotational $S_0$-branches are broadened, $\sim$30~cm$^{-1}$ FWHM. This behavior continues into solid phase I, where both the $S_0$- and $Q_0$-branches exhibit a moderate pressure dependence in frequency, unlike the $E_{2g}$ phonon, but undergo significant pressure-induced broadening increasing by more than one order of magnitude over the investigated pressure range. This broadening is attributed to enhanced anisotropic interactions at higher pressures, which further lift the $m_J$ degeneracy, alongside the hindering of rotations that reduce $J$ as an effective quantum number \cite{Cooke2020, Pena-Alvarez2020, Meier2020}. 

At temperatures below 65~K, hydrogen consists almost entirely of the ground para-H$_2$ ($J=0$) and ortho-H$_2$ ($J=1$) states with the equilibrium ortho-H\textsubscript{2} concentration varying from $40\rightarrow0\%$ as temperature approaches 0~K. Figs. 1 and 2 demonstrate that as the thermal population of the $J=1$ state decreases, the intensities of the $S_0(1)$ and $Q_0$ bands diminish. This relationship is more clearly observed during the ortho-para conversion process shown in Fig. 2, which reveals a strong temporal correlation between the integral intensities of the $Q_0$ and $S_0(1)$ branches at 17~K and 29 GPa. The inset to Fig. 2(b), shows that the $Q_0$ and $S_0(1)$-branch intensities decrease by factors of approximately 5 and 2.5, respectively, over 836 minutes as the system approaches its equilibrium ortho-para concentration. Meanwhile, we see an appreciable 40$\%$ gain in intensity of the $S_0(0)$-branch resulting from an increased population of the $J=0$ state. This demonstrates that below 65~K, the $Q_0$ branch arises entirely from ortho-H$_2$ reorientational transitions within the $J=1$ manifold with no contribution from para-H$_2$ as its $J=0$ ground state is not degenerate. 

The behavioral similarities between the rotational $S_0$-branches and the reorientational $Q_0$-branches persist into phases II and III, structures defined by their quantum and classical orientational ordering, respectively \cite{Mazin1997}. The transition to phase II at 69~K at 100 GPa, is characterized by the disappearance of distinct spectral features, which are replaced by a broad continuum of Raman intensity extending up to 1200~cm$^{-1}$. As shown in End Matter Fig. 6, the minimum between the $Q_0$ and the $S_0$-branches near 200~cm$^{-1}$ vanishes between 91 and 112~GPa. Interestingly, the lack of spectral structure is typically an indicator of a highly disordered state, which in turn presents a paradox, given that this phase has long been proposed to exhibit quantum orientational ordering of the molecular angular momenta\cite{Mazin1997, Mao1994}. At higher pressures, $>$150~GPa and 69~K, the former $S_0(J)$ rotational bands are quenched, transforming into the optical phonons that mark the onset of phase III \cite{Goncharov1998, Goncharov2001_PNAS}, where molecular rotation is inhibited \cite{Mao1994}. A parallel modification also occurs in the $Q_0$-branches, whose broad continuum collapses into two distinct optical phonons below 200~cm$^{-1}$ which are also characteristic of phase III \cite{Goncharov1998}. By correlating excitations across all three phases \cite{Mazin1997}, this observed progression could provide crucial insight for constraining structural models of the currently structurally ambiguous phases of hydrogen, phases II and III \cite{Akahama2017, Loubeyre1996}.

\begin{figure}
    \centering
    \includegraphics{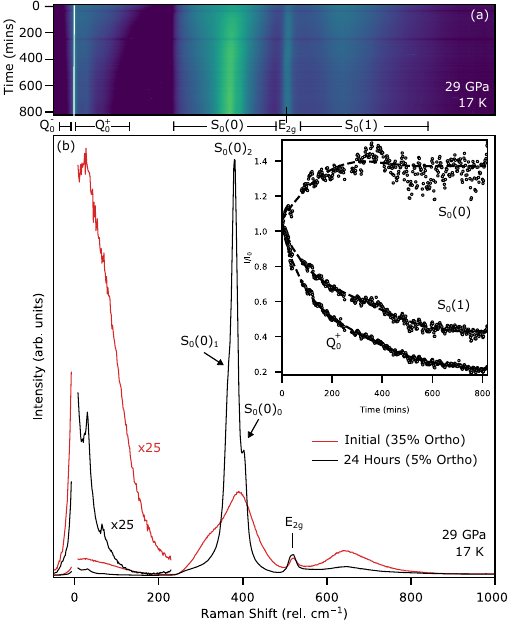}
    \caption{(a) Raman logarithmic intensity heat map as a function of time measured at 17~K and 29~GPa. (b) Raw Raman measurements, normalized to the unstressed diamond edge, taken upon reaching the base temperature (red trace) and 24 hours later (black trace). (b) Inset shows the normalized integrated intensity over time for the $Q_0^+$, $S_0(0)$ and $S_0(1)$ branches shown in panels (a,b).}
    \label{Fig2}
\end{figure}

In our measurements at 17~K and 29~GPa, the ortho-H$_2$ concentration was reduced to approximately 5$\%$ (See End Matter for determination of ortho-para concentration)\cite{Eggert1999}, effectively rendering ortho-H$_2$ to an impurity level within a bulk para-H$_2$ crystal. This high-purity environment is evidenced by the better resolved orientational fine structure of the $S_0(0)$ peak, as seen by comparing the initial (2 hours at 17~K) and final spectra (24 hours at 17~K) in Fig. 2(b). This peak exhibits the characteristic triplet components—$S_0(0)_{|m|}$ for $|m| = 0$, 1, and 2—arising from the partially lifted degeneracy of the $J=2$ rotational level \cite{Goncharov2001_PRB}. Concurrently with the resolution of the $S_0(0)$ triplet, relatively weak but distinct excitations emerge within the formerly broad $Q_0$-branch below 100~cm$^{-1}$. In a subsequent, higher-purity sample (2$\%$ ortho-H$_2$) produced by maintaining the sample at 4.4~K for more than 60 hours, these excitations become even more pronounced as seen in Fig. 3, revealing at least seven partially resolved components below 100~cm$^{-1}$, which we label P$_{1-7}$, and were observed to be almost three orders of magnitude weaker than the most prominent $S_0(0)_2$-excitation.

\begin{figure}
    \centering
    \includegraphics{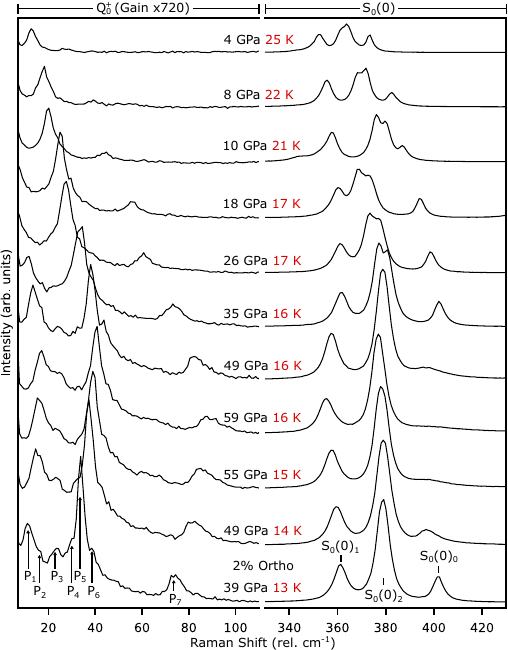}
    \caption{Selected BEP corrected Raman spectra, excited by a 532~nm laser, of the $Q_0$-branch and the $S_0(0)$-branch for a compression (39-59~GPa) followed by decompression (59-4~GPa) of a 2$\%$ ortho-H$_2$ concentration sample. Measurements were performed over the course of several hours while the temperature as temperature increased from 13$\rightarrow$25~K due to the exhaustion of L-He.}
    \label{Fig3}
\end{figure}

The pressure evolution of the $Q_0$ fine-structure frequencies, shown in Fig. 4, provides critical insight into their assignment and origin. The number of observed components and their splitting are consistent with the expected behavior for the $J=1$ manifold, where the sub-level structure depends strongly on the number of $nn$ ortho-H$_2$ molecules surrounding a given ortho-H$_2$ molecule ($nn_{ortho}$). For instance, for $nn_{ortho}=0$ we should only observe two split reorientational sublevels, while if $nn_{ortho}=1$ the total angular momentum quantum number of the pair will be 0, 1, or 2 leading up to 9 split reorientational sub-levels \cite{Hardy1975, Hardy1977, Silvera1980}. Where $nn_{ortho}>1$ the picture rapidly becomes more complex, partially contributing to the broad distribution we find at higher-temperature higher-ortho-H$_2$ concentrations, before expecting to finally simplify for a pure ortho-H$_2$ lattice. In our sample, assuming a perfect random distribution of ortho-H$_2$ molecular impurities (2$\%$ conc.), the resulting fractions are 78$\%$ isolated ($nn_{ortho}=0$), 19$\%$ existing as pairs ($nn_{ortho}=1$), and the remaining 3$\%$ in higher-order configurations (e.g., triplets or quadruplets) (for calculation details see equation (1) in ref. \cite{Loubeyre1992}).

The general hardening of these peaks with pressure, as shown in Fig. 4, highlights the increasing intermolecular anisotropic interactions. A global fit (see End Matter for details) to the pressure dependence of P$_{2-7}$ show they share a common evolution. Extrapolation of these fits to ambient pressure aligns closely with the nine distinct values between 2.1 and 2.9~cm$^{-1}$ (with a manifold predicted to span over 6~cm$^{-1}$ \cite{Silvera1980, Hardy1977}) observed via microwave absorption spectroscopy in solid para-H$_2$ at 1.2~K, which are assigned to reorientational transitions for pairs of $J=1$ ortho-H$_2$ impurities ($nn_{ortho}=1$). In contrast, the linear pressure dependence of the P$_1$ peak suggests a different origin, potentially from isolated ortho-H$_2$ molecules ($nn_{ortho}=0$) \cite{Hardy1975, Hardy1977, Silvera1980}. A similar linear pressure dependence has been reported for isolated hydrogen impurities in a solid neon matrix \cite{Loubeyre1992}, supporting this interpretation. Furthermore, the extrapolation of the linear pressure dependence of P$_1$ to ambient pressure approaches that of the experimentally observed $nn_{ortho}$=0 value, 0.007~cm$^{-1}$ (blue circle in Fig. 4), determined through the fine structure of the infrared $Q_3(0)$ overtone ($\nu=3\leftarrow0$, $J=0\leftarrow0$) \cite{Dickson1996}. 

We report low-frequency Raman excitations, below 250~cm$^{-1}$, in hydrogen across a broad pressure-temperature range, 2-174~GPa and 13-300~K. These peaks are interpreted as transitions among the orientational sub-levels, $m_J$, which are shifted from the Rayleigh line by anisotropic intermolecular interactions. These features constitute a $Q_0$-branch, where the rotational quantum number $J$ is preserved, indicating a quantum reorientational origin. This assignment is supported by the branch's behavior in phase I, which mirrors the rotational $S_0$-branch under compression, as well as the subsequent quenching into phonons in phase III. At temperatures above 65~K, where samples are equilibrated with high ortho-H$_2$ concentration, the $Q_0$-branch exhibits a Boltzmann-like distribution generated by ortho-para configurational disorder and the thermal population of higher $J$-states. Upon cooling below ~65~K, as the system approaches its equilibrium ortho-para concentration and only the $J=0$ and $J=1$ states are thermally populated, the integrated intensity of the $Q_0$-branch exhibits a clear correlation with that of the $S_0(1)$ mode. This correlation strongly suggests the process originates from the $m_J$ degeneracy of the $J$=1 manifold, as the non-degenerate $J=0$ manifold cannot contribute to this band. Finally, a fine structure is observed at low temperatures, less than 20~K, and is attributed to transitions from isolated (P$_1$) and ortho-H$_2$ pairs (P$_{2-7}$). Such transitions have previously been observed only at ambient pressure \cite{Silvera1972, Hardy1975, Dickson1996}; their detection at high pressure provides a sensitive probe of the intermolecular potential in solid H$_2$. Our observations complete the high-pressure data set for the Raman spectrum of the archetypal coupled quantum rotor, solid hydrogen, providing new stimuli for theoretical investigations of anisotropic intermolecular interactions.

\begin{figure}
    \centering
    \includegraphics{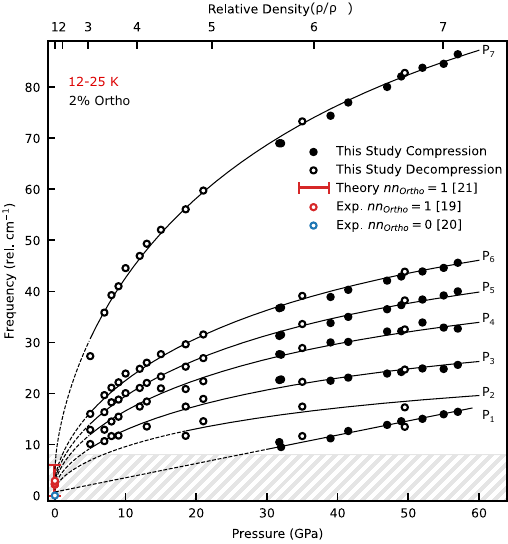}
    \caption{Pressure dependency of low frequency excitations, P$_{1-7}$ (denoted in Fig \ref{Fig3}), from parahydrogen containing $2\%$ ortho-H$_2$ at 13-25~K. Black solid and empty circles denote measurements carried out on compression and decompression, respectively. The solid black traces are fits to a linear function for P$_1$ and a global stretched exponential function for P$_{2-7}$ (see End Matter for functional details), with extrapolations to ambient pressure shown as dashed traces. The empty red circles are taken from 9 ortho-H$_2$ pair excitations reported from microwave measurements on hydrogen at 1.2~K and ambient pressure \cite{Hardy1975} and the red bar represents the theoretical frequency range of the manifold \cite{Harris1977, Silvera1980}. The empty blue circle is the experimentally determined value of an isolated ortho-H$_2$ molecule in parahydrogen from infrared measurements at ambient pressure and low temperatures \cite{Dickson1996}. The gray hatched region represents frequencies which are inaccessible due to the width of the notch filters used $\pm8$~cm$^{-1}$.}
    \label{Fig4}
\end{figure}

\section*{Data Availability Statement}
The data that support the findings of this study are available on request from the authors.

\section*{acknowledgments}
P. D-S. acknowledges the support of the National Natural Science Foundation of China (NSFC) reference code W2532012. M.S. acknowledges HPSTAR for their support as a visiting scientist, during which this research was conducted.

\bibliography{Bib}

\clearpage
\section*{End Matter}
\subsection{Methods}
For this study, we conducted a total of 6 independent experiments, using symmetric diamond anvil cells (DACs) to generate the high pressures, equipped with Type-IIas diamonds with culets 200-50~$\mu$m in diameter and a Boehler-Almax (BA) design allowing for an opening of 70$^{\circ}$. Rhenium gaskets were prepared from 250~$\mu$m foils by pre-indenting them to a thickness of one-fifth the culet diameter and laser-machining a hole two-thirds the culet size in diameter. High-purity Hydrogen (99.99$\%$ $min.$) was gas-loaded into the diamond anvil cell at 1.7~kbar. Throughout the experiment the applied load was generated using pressurized helium gas and a metallic-foil membrane. Pressure was primarily monitored via ruby fluorescence \cite{Shen2020} with associated temperature corrections \cite{Datchi2007} and cross-checked using the Raman-active phonon of the stressed diamond edge \cite{Akahama2006} and the H$_2$ vibron\cite{Howie2013}.

Raman spectroscopy was performed utilizing a micro-focused confocal optical system in a back-scattering geometry. Raman scattering was excited with 532~nm and 660~nm laser wavelengths, producing a spot size of approximately 2~$\mu$m in diameter. To prevent excessive sample heating during measurements, the incident laser power was maintained below 5~mW. The back-scattered light was collected and filtered through ultra-narrow notch filters before being dispersed by a 320~mm aberration corrected spectrometer and imaged using a liquid N$_{2}$-cooled charge-coupled device detector. 1200 and 1800~gr/mm gratings were employed with exposure times of up to 10 minutes in order to reach a suitable spectral resolution and signal-to-noise ratio. Low temperature measurements were carried out using a dry cryostat using either L-N$_{2}$ or L-He as a cryogen, reaching lowest sample temperatures of 65~K and 4.5~K, respectively. 

\begin{figure}
    \centering
    \includegraphics{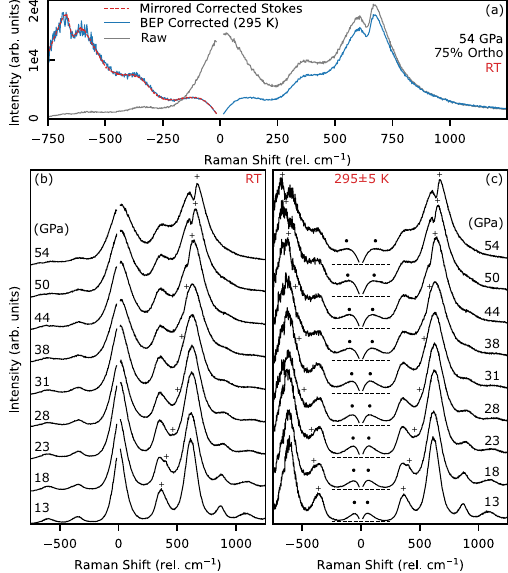}
    \caption{Raman spectra, excited by a 660~nm laser, of solid hydrogen measured upon compression between 13-54 GPa at room temperature. (a) Raw Raman spectrum at 20 GPa (gray), the BEP corrected spectrum (blue) and the mirrored corrected Stokes spectrum (dashed red). (b) Waterfall of the raw Raman spectrum upon compression. (c) Waterfall of the BEP corrected spectra from panel (b), with the average temperature and its error bar (one SD) determined spectroscopically from the BEP corrections. Dashed horizontal lines mark the zero-intensity baselines for each spectrum. In panels (b) and (c), dots and crosses denote the maxima of the $Q_0$-branch and the frequency of the $E_{2g}$ phonon peak, respectively.}
    \label{Fig5}
\end{figure}

\begin{figure}
    \centering
    \includegraphics{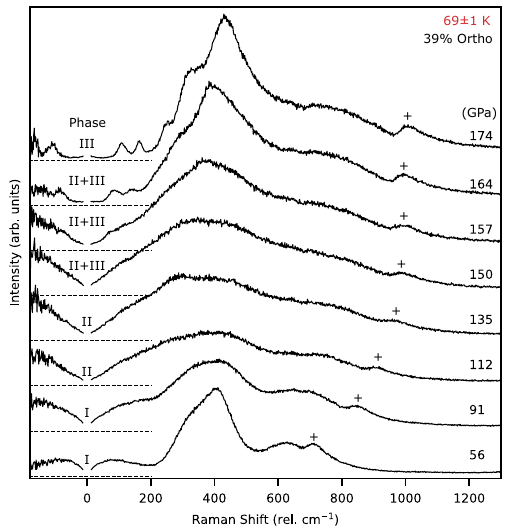}
    \caption{Raman spectra of H$_2$ at 66~K from 56-174~GPa excited by a 532~nm laser, showing the transitions between phases I, II and III. A strong phase coexistence is noted between phases II and III of approximately 25~GPa. The evolution of the $E_{2g}$ phonon (and its equivalent mode in phases II and III \cite{Akahama2017}) is tracked by crosses. Spectral temperatures from fits during isothermal compression average 69$\pm$1~K, with the error corresponding to one SD. Dashed lines indicate the zero-intensity baseline for each spectrum.}
    \label{Fig6}
\end{figure}

\subsection{Bose-Einstein Population Correction}
The raw Raman spectrum is influenced by the Bose-Einstein population factor, $n(\omega, T)$, which introduces a strong, intrinsic temperature and frequency dependence. For a phonon of vibrational frequency $\omega$, this population number at equilibrium is given by:

\begin{equation}
n(\omega, T) = \frac{1}{\exp\left(\frac{\hbar\omega}{k_B T}\right) - 1}.
\end{equation}

The Stokes ($I_s$) and anti-Stokes ($I_{aS}$) intensities are proportional to $n(\omega, T) + 1$ and $n(\omega, T)$, respectively. It is important to note, that this leads to a significant intensity enhancement as $\omega \rightarrow 0$, which can strongly distort low-energy excitation lineshapes, as seen in Figs. \ref{Fig1}(a,c), 2 and \ref{Fig5}(a,b) between -250~cm$^{-1}$ and 250~cm$^{-1}$. Therefore, to recover the underlying Raman spectral profile, which is symmetric around zero for first-order scattering, the spectrum must be divided by this population factor. Expressing the relevant terms in Raman shift, $\Delta\nu$ (which is equal to the phonon frequency $\omega$), this becomes the following BEP correction functions for Stokes, $\chi_{S}(\Delta\nu, T)$, and anti-Stokes, $\chi_{aS}(\Delta\nu, T)$ \cite{Hayes1978}:

\begin{equation}
\chi_{S}(\Delta\nu, T) \propto \frac{I_{S}}{(\nu_i - \Delta\nu)^3(n(\Delta\nu, T) + 1)}
\end{equation}
\begin{equation}
\chi_{aS}(\Delta\nu, T) \propto \frac{I_{aS}}{(\nu_i - \Delta\nu)^3n(\Delta\nu, T)}
\end{equation}

Where $n(\Delta\nu, T) = (\exp[1.439 \Delta\nu / T] - 1)^{-1}$ and $(\nu_i - \Delta\nu)^3$ relates to both fundamental electrodynamics and photoconversion in photon counting systems \cite{Hayes1978}.

We can therefore fit a spectral temperature by minimizing the residuals between the reflected corrected Stokes and corrected anti-Stokes intensities:

\begin{equation}
\min_T f(\Delta\nu, T) = |\chi_{S}(-\Delta\nu, T) - \chi_{aS}(\Delta\nu, T)|
\end{equation}

When applied to high-quality data free from background and fluorescence, this BEP correction with a fitted temperature produces a symmetrized spectrum around the Rayleigh line. The success of this symmetrization unambiguously confirms first-order Raman scattering and yields a temperature consistent with the experimental conditions, as seen in Figs. \ref{Fig1}, \ref{Fig5} and \ref{Fig6}.

\subsection{Determination of Ortho-Para Concentration}
At room temperature, all hydrogen are expected to be at their equilibrium concentration of 75~$\%$ ortho-H$_2$. However, at low temperatures where only the ground states of ortho-H$_2$ ($J=1$) and para-H$_2$ ($J=0$) are populated the relative ortho-concentrations, $x$, were determined using the integrated intensities of the rotational $S_0(0)$-branch ($I_0$) and the $S_0(1)$-branch ($I_1$), following the formula presented by Eggert {\it et al.} \cite{Eggert1999} a simplified equation of what was originally proposed by Silvera {\it et al.} \cite{Silvera1980}:

\begin{equation}
x = \frac{5}{3}\left(\frac{I_0}{I_1} + \frac{5}{3}\right)^{-1}
\end{equation}

\subsection{Fitting of P$_{1-7}$ Pressure Dependencies}

The pressure dependence of the ortho-H$_2$ pair excitation frequencies P$_{2-7}$ were fitted using the global stretched exponential function, equation (6). 

\begin{equation}
\nu(P) = A_i \left( \nu_\infty \left( 1 - \exp\left[-\left(\frac{P}{P_0}\right)^{\alpha}\right] \right) + \nu_0 \right)
\end{equation}

where $A_i$ is an individual scaling factor for each mode (provided in Table I), while $\nu_\infty$ (29.1~cm$^{-1}$), $P_0$ (40.6~GPa), $\alpha$ (0.636), and $\nu_0$ (0.961~cm$^{-1}$) are shared parameters across all peaks. This functional form captures the gradual saturation of the phonon frequencies at high pressure while preserving a smooth evolution at low pressure. 

The P$_1$ mode, suggested to be from isolated ortho-H$_2$ molecules, exhibited instead a distinct linear dependence and was fitted with equation (7) with the parameters $\nu_0=0.28$~cm$^{-1}$ and $\beta=0.71$~cm$^{-1}$GPa$^{-1}$. 

\begin{equation}
\nu(P) = \nu_0 + \beta P
\end{equation}

\begin{table}
\caption{Individual scaling factors ($A_i$) used in equation (6) for the global stretched-exponential fit for ortho-H$_2$ pair excitations (P$_{2-7}$).}
\label{Tab1}
\begin{tabular}{lc}
Excitation & $A_i$ \\
\hline
P$_2$ & 0.89 \\
P$_3$ & 1.20 \\
P$_4$ & 1.55 \\
P$_5$ & 1.82 \\
P$_6$ & 2.10 \\
P$_7$ & 3.97 \\
\end{tabular}
\end{table}

\end{document}